\documentclass[11pt]{article}
\usepackage{graphicx}        
\usepackage{multicol}        
\usepackage[bottom]{footmisc}
\usepackage{bm}
\usepackage{amsmath}
\usepackage{amssymb}
\usepackage{caption}
\usepackage[noadjust]{cite}

 \newtheorem{thm}{Theorem}

 \newtheorem{example}[thm]{Example}
 
 \numberwithin{equation}{section}

\newcommand{\hf}{\frac{1}{2}}

\newcommand{\gvec}[1]{\ensuremath{\mbox{\textbf{\textit{#1}}}}} 
\newcommand{\sgvec}[1]{\ensuremath{\mbox{\textbf{\textit{\small #1}}}}}
\newcommand{\eo}{e_0}
\newcommand{\einf}{e_{\infty}}
\newcommand{\be}{\begin{equation}}
\newcommand{\ee}{\end{equation}}

  \newcommand{\R}{\ensuremath{\mathbb{R}}}

  \newcommand{\Z}{\ensuremath{\mathbb{Z}}}
   
\begin{document}

\centerline{\bf RECIPROCAL SPACE AND CRYSTAL PLANES IN GEOMETRIC ALGEBRA}

\vspace{11pt}

\centerline{ {\bf E.~Hitzer}}

\vspace{11pt}

\centerline{Department of Applied Physics}
\centerline{University of Fukui, Fukui, Japan} 
\centerline{hitzer@mech.u-fukui.ac.jp}

\vspace{11pt}

This contribution discusses the geometry of $k$D crystal cells given by $(k+1)$ points in a projective space $\R^{n+1}$. We show how the concepts of barycentric and fractional (crystallographic) coordinates, reciprocal vectors and dual representation are related (and geometrically interpreted) in the projective geometric algebra $\R_{n+1}$ (see \cite{HG:AL1844}) and in the conformal algebra $\R_{n+1,1}$. The crystallographic notions of $d$-spacing, phase angle (in structure factors), extinction of Bragg reflections, and the interfacial angles of crystal planes are obtained in the same context.

\vspace{7pt}
\noindent
\textbf{1 \hspace{7pt} Crystal planes in geometric algebra of projective space}
\vspace{7pt}

\noindent
We first consider the \textit{barycentric coordinates} of projective geometry and their relationship to the \textit{fractional coordinates} of crystallography. 

In an offset 2D plane a point $x\in\R^{n+1}$ can be represented as a linear combination of three points $a, b, c \in\R^{n+1}$ in general location. (We can set $n=3$, but our results are valid for general $n$.) If the three points are of unit weight, the linear combination becomes an \textit{affine combination}
\be 
  x = \alpha a + \beta b + \gamma c, \quad \alpha + \beta + \gamma = 1, 
  \quad \alpha, \beta, \gamma \in \R. 
\ee 
We compute the coefficients by subtracting $c$ on both sides and wedging with $\gvec{a} = a-c \in\R^n$ and $\gvec{b} = b-c\in\R^n$, respectively
\be 
  \gvec{x} = x-c = \alpha \gvec{a} + \beta \gvec{b} + (\alpha+\beta+\gamma-1)c
  = \alpha \gvec{a} + \beta \gvec{b} \in\R^n.
\ee 
We obtain $\beta$ as a ratio of oriented areas by
\be 
  \gvec{x} \wedge \gvec{a} = \beta \gvec{b}\wedge \gvec{a}
  \Rightarrow 
  \beta =  \frac{\gvec{x} \wedge \gvec{a}}{\gvec{b}\wedge \gvec{a}}
  = -\frac{\gvec{x} \wedge \gvec{a}}{\gvec{a}\wedge \gvec{b}},
  \,\,\,\text{ and similarly }\,\,\,
  \alpha = \frac{\gvec{x} \wedge \gvec{b}}{\gvec{a}\wedge \gvec{b}},
\ee 
and therefore
\be 
  \gamma = 1 - \alpha - \beta 
  = 1- \frac{\gvec{x} \wedge \gvec{b}}{\gvec{a}\wedge \gvec{b}}
     + \frac{\gvec{x} \wedge \gvec{a}}{\gvec{a}\wedge \gvec{b}}
  =  1- \frac{\gvec{x} \wedge (\gvec{b}-\gvec{a})}{\gvec{a}\wedge \gvec{b}}.
\ee 
In the denominator of $\alpha$ and $\beta$ we have the oriented volume (area) $\gvec{a}\wedge \gvec{b} \in \bigwedge^2\R^n$ of a cell (parallelogram spanned by $\gvec{a}$ and $\gvec{b}$).
In the crystallography of 2D crystals, point $c$ is often called the origin of the cell, and the coordinate values $\alpha$ and $\beta$ are called \textit{fractional coordinates}. The \textit{barycentric coordinates} $\alpha, \beta, \gamma$ can be used to interpolate a scalar property $S: \R^n\rightarrow \R$ given at the vertexes $a, b, c$ to a value at $x$: $S_x = \alpha S_a + \beta S_b + \gamma S_c$.

We further observe, that the vectors 
\be 
  \gvec{a}' = \gvec{b}/(\gvec{a}\wedge \gvec{b}), \quad
  \gvec{b}' = -\gvec{a}/(\gvec{a}\wedge \gvec{b})
\ee 
are also called \textit{reciprocal vectors} in crystallography. They have the property that
\be 
  \gvec{a}\rfloor \gvec{a}' 
  = \gvec{a}\rfloor (\gvec{b}(\gvec{a}\wedge \gvec{b})^{-1})
  = (\gvec{a}\wedge \gvec{b})(\gvec{a}\wedge \gvec{b})^{-1} = 1,
\ee 
by applying $(A\wedge B)\rfloor C = A\rfloor (B\rfloor C),\,\, \forall A,B,C \in \R_{n+1}$, for the second equality. Similarly
\be 
  \gvec{b}\rfloor \gvec{b}' =1, 
  \quad
  \gvec{a}\rfloor \gvec{b}' 
  = \gvec{a}\rfloor [\gvec{a}(\gvec{a}\wedge \gvec{b})^{-1}]
  = (\gvec{a}\wedge \gvec{a})(\gvec{a}\wedge \gvec{b})^{-1} = 0,
  \quad
  \gvec{b}\rfloor \gvec{a}' =0.
\ee 
Since, by applying again $(A\wedge B)\rfloor C = A\rfloor (B\rfloor C)$, we can conversely rewrite the coefficient equations as
\be 
  \alpha = \frac{\gvec{x} \wedge \gvec{b}}{\gvec{a}\wedge \gvec{b}}
         = \gvec{x} \rfloor [\gvec{b}/({\gvec{a}\wedge \gvec{b}})]
         = \gvec{x} \rfloor \gvec{a}',
  \,\,\,\text{ and similarly }\,\,\,
  \beta = \gvec{x} \rfloor \gvec{b}',
\ee 
where we observe the familiar role of reciprocal vectors in crystallography.
Note that geometrically the inverse of $\gvec{a}'$ is the rejection of $\gvec{a}$ from $\gvec{b}$
\be 
  \gvec{a}'^{-1} = (\gvec{a}\wedge \gvec{b})\gvec{b}^{-1} = P^{\perp}_{\sgvec{b}}(\gvec{a}),
\ee 
which can be interpreted as the perpendicular distance vector of $a$ from the line $c\wedge b$.
And likewise
\be 
  \gvec{b}'^{-1} = (\gvec{b}\wedge \gvec{a})\gvec{a}^{-1} = P^{\perp}_{\sgvec{a}}(\gvec{b}),
\ee 
the perpendicular distance vector of $b$ from the line $c\wedge a$.

If we add the two reciprocal vectors we obtain 
\be 
  \gvec{a}' + \gvec{b}' 
  = \gvec{b}/(\gvec{a}\wedge \gvec{b})
   -\gvec{a}/(\gvec{a}\wedge \gvec{b})
  = (\gvec{b}-\gvec{a})/(\gvec{a} \wedge \gvec{b})
  = \gvec{s}^{-1},
  \label{eq:2DaddRcVec}
\ee 
i.e. the inverse of the support vector $\gvec{s}\in\R^n$ (relative to point $c$) of the line $a\wedge b = (c+\gvec{s})(\gvec{b}-\gvec{a})$ with direction vector $\gvec{b}-\gvec{a}$.

In Grassmann algebra and in Clifford's geometric algebra the subspace spanned by two linearly independent vectors is given by their outer product. And in projective geometric algebra indeed the outer product of two points (interpreted as outer product null space [OPNS]) spans the line through these two points, including its offset from the origin $e_0$
\be 
  \label{eq:Lbiv}
  p \wedge q 
  = e_0\wedge (\gvec{q}-\gvec{p}) + \gvec{p}\wedge \gvec{q}
  = e_0\wedge \gvec{a} + \gvec{M},
\ee 
where we used unit points (scalar multiples would span the same subspace). 
For any point $x=\alpha p + \beta q, \forall \alpha,\beta \in \R$ on the line has zero outer product with $p \wedge q$
\be 
  x \wedge (p\wedge q) = 0 \,\,\,
  \Leftrightarrow \,\,\,
  x = \alpha p + \beta q, \,\,\, \alpha,\beta \in \R.
\ee 
We recognize $\gvec{a}=\gvec{q}-\gvec{p}\in\R^n$ as the direction vector of the line. Reshaping $\gvec{M}=\gvec{p}\wedge \gvec{q}$ by Cramer's rule to rectangular shape
\be 
  \label{eq:LMoment}
  \gvec{M}
  = \gvec{p}\wedge (\gvec{q}-\gvec{p})
  = \gvec{r} (\gvec{q}-\gvec{p})\,\,\,
  \Leftrightarrow
  \,\,\,
  \gvec{r} = (\gvec{p}\wedge \gvec{q}) (\gvec{q}-\gvec{p})^{-1}
           = \gvec{M}\gvec{a}^{-1},
\ee 
we find it to be the \textit{moment} (bivector) $\gvec{M}=\gvec{r}\gvec{a}$ of the line, i.e. the geometric product of its distance vector $\gvec{r}\in\R^n$ from the origin, times $\gvec{a}$. $\gvec{r}$ is also called (perpendicular) support vector (relative to the origin $e_0$) with projective point representation 
\be 
  r = e_0 + \gvec{r}.
\ee 
An alternative specification is the point-direction representation (or line-bound vector from an affine perspective) of the line space bivector
\be 
  p\wedge q = p \wedge (p-q) = p \wedge \gvec{a},
\ee 
where $p$ is a point on the line and $\gvec{a}$ the direction vector.

The above example of an offset 2D plane spanned by 3 points can easily be generalized to barycentric coordinates in a 3D subspace spanned by 4 points $a,b,c,d$. If these points are unit-weight points, then an affine combination represents every other point 
$x=\alpha a + \beta b + \gamma c + \delta d$. With analogous definitions $\gvec{x} = x-d$, etc., to the plane case the first three coefficients $\alpha, \beta, \gamma$ will be the usual \textit{fractional coordinates} of crystallography for a 3D crystal cell with origin $d$. 

The result is that in the 3D case the first three barycentric coordinates of $x$, corresponding to fractional coordinates of crystallography, are given by
\be 
  \alpha = \frac{\gvec{x} \wedge \gvec{b} \wedge \gvec{c}}{\gvec{a}\wedge \gvec{b}\wedge \gvec{c}}, \quad
  \beta = \frac{\gvec{a} \wedge \gvec{x} \wedge \gvec{c}}{\gvec{a}\wedge \gvec{b}\wedge \gvec{c}}, \quad 
  \gamma = \frac{\gvec{a} \wedge \gvec{b} \wedge \gvec{x}}{\gvec{a}\wedge \gvec{b}\wedge \gvec{c}},
\ee 
where $\gvec{a} = a - d, \gvec{b}=b-d, \gvec{c}= c-d$.

The \textit{reciprocal vectors} of a 3D cell with cell vectors $\gvec{a}, \gvec{b}, \gvec{c}$ are 
\be
  \gvec{a}' = \frac{\gvec{b} \wedge \gvec{c}}{\gvec{a}\wedge \gvec{b}\wedge \gvec{c}}, \quad 
  \gvec{b}' = \frac{\gvec{c} \wedge \gvec{a}}{\gvec{a}\wedge \gvec{b}\wedge \gvec{c}}, \quad
  \gvec{c}' = \frac{\gvec{a} \wedge \gvec{b}}{\gvec{a}\wedge \gvec{b}\wedge \gvec{c}},
\ee 
where the denominator $\gvec{a}\wedge \gvec{b}\wedge \gvec{c}$ is the oriented 3-volume of the cell, and the numerator is obtained by removing the vector, whose reciprocal vector is to be defined. 
The inverse reciprocal vectors are heights of corresponding points over a side face of the parallelepiped cell
\be 
  \gvec{a}'^{-1}=P_{\sgvec{b}\wedge\sgvec{c}}^{\perp}(\gvec{a}), 
  \quad
  \gvec{b}'^{-1}=P_{\sgvec{c}\wedge\sgvec{a}}^{\perp}(\gvec{b}),
  \quad
  \gvec{c}'^{-1}=P_{\sgvec{a}\wedge\sgvec{b}}^{\perp}(\gvec{c}),
\ee 
i.e. the height vectors of $a$ over side plane $b \wedge c \wedge d$, 
of $b$ over $c \wedge a \wedge d$, and $c$ over 
$a \wedge b \wedge d$, respectively.
The reciprocal vectors fulfill 
\be 
  \gvec{a}\rfloor \gvec{a}' 
  = \gvec{b}\rfloor \gvec{b}' 
  = \gvec{c}\rfloor \gvec{c}' =1, \quad
  \text{ while all other products vanish } \quad
  \gvec{b}\rfloor \gvec{a}' = \gvec{c}\rfloor \gvec{a}' = 0, \,\ldots
\ee 

We further observe that adding the three reciprocal vectors gives the inverse of the support vector (relative to point $d$) of the plane 
$a\wedge b\wedge c = s\gvec{A}, \,\,s = (d+\gvec{s}) \in \R^{n+1}$, 
with direction bivector $\gvec{A}$ and moment trivector $\gvec{M}$
\be 
  \gvec{s}^{-1} 
  = \gvec{a}'+\gvec{b}'+\gvec{c}'
  = \frac{\gvec{b} \wedge \gvec{c}+\gvec{c} \wedge \gvec{a}+\gvec{a} \wedge \gvec{b}}{\gvec{a}\wedge \gvec{b}\wedge \gvec{c}}
  = \gvec{A}/\gvec{M}. 
\ee  

The concept of affine combination of $k+1$ (unit-weight) points $a_1, \ldots, a_k, a$ to determine any other point $x$ in the offset $k$D subspace thus spanned leads to the corresponding barycentric and fractional coordinates, as well as the reciprocal vectors ($\check{\gvec{a}}_l$ means to omit $\gvec{a}_l=a_l-a$)
\be 
  \gvec{a}{'}_l= (-1)^{l-1}(\gvec{a}_1\wedge \ldots \check{\gvec{a}}_l\ldots\wedge \gvec{a}_k)/(\gvec{a}_1\wedge \ldots \wedge \gvec{a}_k),
\ee
of a $k$D parallelepiped cell of crystallography. The inverse of each reciprocal vector
\be
 \gvec{a}{'}_l^{-1} = P_{(-1)^{l-1}\sgvec{a}_1\wedge \ldots \check{\sgvec{a}}_l\ldots\wedge \sgvec{a}_k}^{\perp}(\gvec{a}_l)
\ee
is again the rejection of $\gvec{a}_l$ from the $(k-1)$D side face 
$\gvec{a}_1\wedge \ldots \check{\gvec{a}}_l\ldots\wedge \gvec{a}_k$, i.e. the
height vector of point $a_l$ over the side face $(k-1)$D plane ${a}_1\wedge \ldots \check{{a}}_l\ldots\wedge {a}_k\wedge a$.
Vectors and reciprocal vectors are related by
\be 
  \gvec{a}_m\rfloor \gvec{a}{'}_l = \delta_{m,l}, \quad 1\leq m,l \leq k,
\ee
where $\delta_{m,l}= 1$ for $m=l$, and $\delta_{m,l}= 0$ for $m\neq l$.
Any vector $x$ in the offset $k$D subspace spanned by $a_1, \ldots a_k, a$ can be represented as
\begin{align} 
  \hspace{-2mm}x = \hspace{-1mm}\sum_{l=1}^k (\gvec{x}\rfloor \gvec{a}{'}_l) a_l 
      + (1-\hspace{-1mm}\sum_{l=1}^k \gvec{x}\rfloor \gvec{a}{'}_l) a
    = a+\hspace{-1mm}\sum_{l=1}^k (\gvec{x}\rfloor \gvec{a}{'}_l) (a_l-a)
  \Leftrightarrow 
  \gvec{x} = x-a = \hspace{-1mm}\sum_{l=1}^k (\gvec{x}\rfloor \gvec{a}{'}_l) \gvec{a}_l
  \label{eq:PoffsSubspParEq}
\end{align}
The first $k$ barycentric coordinates $\gvec{x}\rfloor \gvec{a}{'}_l, 1\leq l \leq k$ 
correspond thus to the fractional coordinates of a $k$D crystallographic cell, projectively embedded in $\R^{n+1}$. Equation \eqref{eq:PoffsSubspParEq} represents a \textit{parametric equation} of a $k$D offset subspace in projective geometric algebra, to be compared with the $(k+1)$-blade representation $a_1 \wedge \ldots \wedge a_k \wedge a$.

The support vector of the $(k-1)$D offset subspace $\Pi_{k-1}=a_1\wedge \ldots \wedge a_k$ relative to point $a$ is given by
\be 
  \gvec{d}^{-1} = \sum_{l=1}^k \gvec{a}{'}_l.
\ee 

If the offset subspace is a hyperplane, i.e. $k=n$, then its \textit{dual} representation (as inner product null space [IPNS]) is given by the point 
\be 
  \pi =\Pi_{n-1}(\gvec{I}a)^{-1}= a-\gvec{d}^{-1}=a-\sum_{l=1}^n \gvec{a}{'}_l \in \R^{n+1}.
\ee 
The previously mentioned OPNS representation and the dual IPNS representation are directly related by the duality operation (multiplication by the inverse pseudoscalar of $\R^{n+1}$, $\gvec{I}$ is the pseudoscalar of $\R^n$) 
\begin{align}
  x \rfloor \pi 
  = x \rfloor [\Pi_{n-1}(\gvec{I}a)^{-1}]
  = (x \wedge \Pi_{n-1})(\gvec{I}a)^{-1},
\end{align}
which holds again because of  $(A\wedge B)\rfloor C = A\rfloor (B\rfloor C)$.

Therefore if the points $a_1, \ldots a_n, a$ represent an $n$D crystal cell, with $a$ as the origin of the cell, then any hyperplane of the space lattice formed by applying translations $T_{\sgvec{a}_l}[\,], 1\leq l \leq n$, can be dually represented by points $\pi$ (setting $a=e_0$ for simplicity)
\be 
  \pi_{(h_1, \ldots, h_n)} 
  = e_0-\gvec{d}^{-1}_{(h_1, \ldots, h_n)} 
  = e_0-\sum_{l=1}^n h_l \gvec{a}{'}_l,
  \quad 
  \forall\, h_l\in \Z, 1\leq l \leq n.
\ee 
For $n=3$ the (relatively prime) integer coefficients are usually called \textit{Miller indexes}
\be 
  \pi_{(h k l)} 
  = e_0-\gvec{d}^{-1}_{(h k l)} 
  = e_0- (h \gvec{a}{'}_1 + k \gvec{a}{'}_2+l \gvec{a}{'}_3),
  \quad 
  \forall\, h,k,l \in \Z.
\ee 

\begin{example}[Plane with $(hkl)=(1,3,2)$]
Assume a plane with Miller indexes $(hkl)=(1,3,2)$. Then we can immediately write down the dual point form of the plane
\be 
  \pi_{(1, 3, 2)} 
  = e_0-\gvec{d}^{-1}_{(1, 3, 2)} 
  = e_0- (1 \gvec{a}{'} + 3 \gvec{b}{'}+2 \gvec{c}{'}).
\ee 
We can compute the distance vector between two neighboring planes as
\be 
  -\gvec{d}^{-1}_{(1, 3, 2)} 
  = e_0\rfloor(e_0\wedge \pi_{(1, 3, 2)})
  \Rightarrow 
  \gvec{d}_{(1, 3, 2)} = -[e_0\rfloor(e_0\wedge \pi_{(1, 3, 2)})]^{-1}
\ee 
\end{example}
The socalled $d$-\textit{spacing} of two neighboring planes is given by the length $|\gvec{d}|$ of the support vector $\gvec{d}$
\be 
  d_{hkl} 
  = |\gvec{d}_{(hkl)}| = |e_0\rfloor(e_0\wedge \pi_{(hkl)})|^{-1}. 
\ee

\vspace{7pt}
\noindent
\textbf{2 \hspace{7pt} Crystal planes in conformal geometric algebra}
\vspace{7pt}

\noindent
The conformal
model 
of Euclidean space (in the GA of $\mathbb{R}^{3+1,1}$), 
which adds two null-vector dimensions for 
the origin $\gvec{e}_0$ and infinity $\gvec{e}_{\infty}$ such that
\begin{align}
  X = \gvec{x} + \frac{1}{2}\gvec{x}^2{e}_{\infty}+{e}_0, 
\qquad
  {e}_0^2 = {e}_{\infty}^2=X^2=0,
\qquad
  X\rfloor {e}_{\infty} = -1.
\end{align}
The $+{e}_0$ term integrates projective geometry,
and the $+ \frac{1}{2}\gvec{x}^2{e}_{\infty}$ term
ensures $X^2=0$.
The inner product of two conformal points gives their Euclidean distance and 
therefore (in IPNS) a plane $\mu$ equidistant from two points $A,B$ as
\begin{align}
  X \rfloor A = -\frac{1}{2}(\gvec{x}-\gvec{a})^2 \,\,
  &\Rightarrow \,\,X\rfloor (A-B)=0,
  \\
  \mu=A-B &\propto \gvec{n}+d\,{e}_{\infty},
\end{align}
where $\gvec{n}$ is a unit normal to the plane and $d$ its signed scalar distance 
from the origin. Reflecting at two parallel planes $\mu,\mu^{\prime}$ with 
distance $\gvec{t}/2$ we get the so-called \textit{translator} 
(\textit{transla}tion opera\textit{tor} by $\gvec{t}\,$)
\begin{equation}
  X^{\prime} = \mu^{\prime}\mu\,X\,\mu\mu^{\prime} = T_{\gvec{t}}^{-1} X T_{\gvec{t}},
  \quad T_{\gvec{t}}=1+\frac{1}{2}\gvec{t}{e}_{\infty}.
\end{equation}
Reflection at two non-parallel planes $\mu,\mu^{\prime}$ yields the rotation around
the $\mu,\mu^{\prime}$-intersection line axis by twice the angle subtended by $\mu,\mu^{\prime}$.

Group theoretically the conformal group $C(3)$ is isomorphic to $O(4,1)$ and the
Euclidean group $E(3)$ is the subgroup of $O(4,1)$ leaving infinity 
${e}_{\infty}$ invariant.
Now general translations and rotations are represented by geometric products
of vectors. For an application of these concepts to interactive crystal symmetry visualization see \cite{EH:Grassmann200,PH:SGV}.

A 2D crystal plane $(hkl)$ through three conformal points
\begin{align} 
  A_h = \eo + \gvec{a}/h + \hf \frac{\gvec{a}^2}{h^2} \einf, \,\,\,
  B_k = \eo + \gvec{b}/k + \hf \frac{\gvec{b}^2}{k^2} \einf, \,\,\, 
  C_l = \eo + \gvec{c}/l + \hf \frac{\gvec{c}^2}{l^2} \einf, 
\end{align} 
is given (in OPNS) by the conformal 4-blade
\be 
  \Pi_{(hkl)} = A_h\wedge  B_k \wedge C_l \wedge \einf.
\ee 
The reciprocal (dual) conformal vector represntation of the $(hkl)$ plane (IPNS) is
\be 
  \pi_{(hkl)}= \Pi_{(hkl)}^{\ast} = \gvec{d}_{hkl}^{-1} + \einf. 
\ee 
The meet of any point $P=\eo + \gvec{p} + \hf \gvec{p}^2 \einf$ with the plane $\Pi_{(hkl)}$ gives the distance of $P$ from $\Pi_{(hkl)}$ in units of the $d$-spacing of $\Pi_{(hkl)}$
\be 
  P \vee \Pi_{(hkl)} 
  = \Pi_{(hkl)}^{\ast} \lfloor P
  = (\gvec{d}_{hkl}^{-1} + \einf)\lfloor (\eo + \gvec{p} + \hf \gvec{p}^2 \einf)
  = \gvec{d}_{hkl}^{-1}\ast \gvec{p} -1.
\ee 
The meet $P \vee \Pi_{(hkl)} $ allows therefore to directly compute the \textit{phase angle} in the \textit{structure factor} $F(hkl)$ of an atom at point $P$ in a crystal cell as
\be 
  \phi = 2\pi P \vee \Pi_{(hkl)}.
\ee  

Selecting a general position vector $P$ with 
$\gvec{p}=x\gvec{e}_1+y\gvec{e}_2+z\gvec{e}_3$
and subjecting it to a space group symmetry operation given by a conformal versor $V: P \mapsto P' = Ad_V(P) = \widehat{V}^{-1} P V$,
allows to compute the phase angle of the symmetrical position $P'$
\be 
  \phi' = 2\pi P' \vee \Pi_{(hkl)}.
\ee  
This may lead to extinction of the $hkl$ Bragg reflection for 
\be 
  \phi' - \phi 
  = 2\pi (P'-P)\vee \Pi_{(hkl)} 
  = (2n+1) \pi, \,\,n \in \Z.
\ee 

The \textit{interfacial angle} $\theta$ between two crystal planes $(h_1k_1l_1)$ and $(h_2k_2l_2)$ given by their reciprocal representation vectors
\begin{align}
  \pi_1 = \pi_{h_1k_1l_1}= \gvec{d}_{h_1k_1l_1}^{-1} + \einf = \gvec{d}_1^{-1}+ \einf, \,\,\,
  \pi_2 = \pi_{h_2k_2l_2}= \gvec{d}_{h_2k_2l_2}^{-1} + \einf = \gvec{d}_2^{-1}+ \einf,
\end{align}
 is defined as
\be 
  \cos \theta 
  = \frac{\Pi_1 \ast  \Pi_2}{|\Pi_1| |\Pi_2|}
  = \frac{\pi_1 \ast \pi_2}{|\pi_1| |\pi_2|}
  = \frac{(\gvec{d}_{1}^{-1}+\einf)\ast(\gvec{d}_{2}^{-1}
    +\einf)}{|\gvec{d}_{1}|^{-1} |\gvec{d}_{2}|^{-1}}
  = \frac{\gvec{d}_{1}\ast\gvec{d}_{2}}{|\gvec{d}_{1}| |\gvec{d}_{2}|},
\ee 
since $\gvec{d}_{j}^{-1}= \gvec{d}_{j}/|\gvec{d}_{j}|^2$, $\gvec{d}_{j}\ast \einf=0$, $j=1,2$, and because the factors $|\gvec{d}_{j}|^2$, $j=1,2$, in the numerator and denominator cancel out.

By instead using unit norm reciprocal representation vectors, we can use the counter $n \in \Z$ in
\be 
  \pi_{(hkl)} = \gvec{n}_{(hkl)}+n \,d_{(hkl)}\,\einf, 
  \quad \pi_{(hkl)}^2 = \gvec{n}_{(hkl)}^2=1,
\ee 
to faithfully represent every single plane in the family of crystal planes given by the reciprocal vector $\gvec{d}_{(hkl)}^{-1} = \gvec{n}_{(hkl)}/d_{(hkl)} \in \R^n$. For $n=0$ the plane $\pi_{(hkl)} = \gvec{n}_{(hkl)}$ includes the origin.



\begin{thebibliography}{00}

\bibitem{MMJ:FCCA} 
  M.M. Julian, 
  \textit{Foundations of Crystallography with Computer Applications}, 
  CRC Press, Taylor \& Francis Group, Boca Raton, 2008.

\bibitem{HG:AL1844}
  H. Grassmann, edited by F. Engel, 
  \textit{Die Ausdehnungslehre von 1844 und die Geom. Anal.},
  vol. 1, part 1, Teubner, Leipzig, 1894.

\bibitem{EH:Grassmann200}
  E. Hitzer, 
  \textit{New views of crystal symmetry guided by profound admiration of the extraordinary works of Grassmann and Clifford}, 
  accepted for: Proc. of Grassmann Bicentennial Conf., Potsdam (DE), Szczecin (PL), Sep. 2009.

\bibitem{PH:SGV}
C. Perwass, E. Hitzer, (2005) \textit{Space Group Visualizer}, www.spacegroup.info (with free demo for space groups 1, 88, 230). The Space Group Visualizer can be purchased from Raytrix GmbH, www.raytrix.de


\end{thebibliography}
\end{document}